\documentclass[epjST]{svjour}

\usepackage{amsmath}
\usepackage{amssymb,amsfonts,latexsym}
\usepackage{bm}
\usepackage[mathcal]{euscript}
\usepackage{graphicx}
\usepackage{epsfig}
\usepackage{color}

\newcommand{\beq}{\begin{equation}}
\newcommand{\eeq}{\end{equation}}

\begin{document}

\title{Comment on Ihle, ``Towards a quantitative kinetic theory of polar active matter''}

\author{Eric Bertin\inst{1,2} \and Hugues Chat\'{e}\inst{3,4} \and Francesco Ginelli\inst{5} \and Guillaume Gr\'egoire\inst{6} \and S\'ebastien L\'eonard\inst{3,6} \and Anton Peshkov\inst{7}}

\institute{Laboratoire Interdisciplinaire de Physique, Universit\'e Joseph Fourier Grenoble, CNRS UMR 5588, BP 87, 38402 Saint-Martin d'H\`eres, France
\and Universit\'e de Lyon, Laboratoire de Physique, ENS Lyon, CNRS, 46 all\'ee d'Italie, 69007 Lyon, France
\and Service de Physique de l'Etat Condens\'e, CNRS URA 2464, CEA-Saclay, 91191 Gif-sur-Yvette, France
\and LPTMC, CNRS UMR 7600, Universit\'e Pierre et Marie Curie, 75252 Paris, France
\and SUPA, Institute for Complex Systems and Mathematical Biology,
King's College, University of Aberdeen, Aberdeen AB24 3UE, United Kingdom
\and Laboratoire Mati\`ere et Syst\`emes Complexes (MSC), Univ.~Paris-Diderot, CNRS UMR 7057, 75205 Paris Cedex 13
\and Physique et M\'ecanique des Milieux H\'et\'erog\`enes, CNRS UMR 7636, Ecole Sup\'erieure de Physique et de Chimie Industrielles, 10 rue Vauquelin, 75005 Paris, France}



\date{\today}

\abstract{A comment on the preprint "Towards a quantitative kinetic theory of polar active matter" by T.~Ihle, arXiv:1401.8056.}
%

\maketitle

In his contribution \cite{IHLE-DD}, Ihle reviews the kinetic theory approach he developed \cite{IHLE} to derive hydrodynamic equations for the Vicsek model \cite{VICSEK,CHATE}. He further presents a critical assessment of the Boltzmann approach initiated by some of us \cite{BDG} (denoted below as BDG), and later extended to other Vicsek-type models with different symmetries \cite{RODS-KINETIC,ACTIVE-NEMA} or non-metric interaction range \cite{TOPOKINETIC} (in the latter case, see also the work by Ihle et.al.~\cite{IHLE-TOPO}). Further results on the case of polar particles with ferromagnetic interactions can also be found in \cite{BGL-DD,ANTON_PHD}, as well as in \cite{MCM-polar}.

\medskip
\noindent
{\bf Qualitative vs.~quantitative results.}
In \cite{IHLE-DD}, Ihle claims that the kinetic approach he uses to derive hydrodynamic equations is the only one, among the approaches known in the literature, designed to yield a quantitative agreement with the Vicsek model.
In spirit, we agree with this statement, in the sense that the Boltzmann approach we developed was not aimed to provide a quantitative description of the Vicsek model \cite{BDG}. Rather, the goal of this work was to derive hydrodynamic equations starting from a model different from the Vicsek model (but however in the same class, see below), using several approximations. The key idea is thus to obtain hydrodynamic equations that are on the one hand well-behaved, on the other hand representative of the entire class of self-propelled particles systems with metric ferromagnetic interactions.

In this respect, we believe that Ihle overestimates the importance of the notion of quantitative agreement in the field of active matter modeling, at least in its present state, due to the lack of realistic modeling of the microscopic dynamics (see however \cite{Gautrais,Weber-VPD}). When studying usual fluids, it may be possible to determine with a relatively good accuracy the interactions between molecules. It is thus relevant in this case to try to perform a quantitative coarse-graining procedure and to compare the large-scale theory with experimental results. In contrast, the Vicsek model is, from a quantitative viewpoint, likely to be far from any realistic system of self-propelled particles. Its interest lies in its ability to reproduce emergent flocking patterns, in its computational efficiency, and in its historical role in the development of the field. In other words, it is a relevant representative model of the class of self-propelled particle models with ferromagnetic interactions on a metric range.
Of course, it is a priori better to have a quantitative coarse-graining theory at hand rather than a qualitative one, but the importance of having quantitative results on the Vicsek model should not be overstated, since the Vicsek model is not more relevant than any other model of the same class.

Let us also mention that the notion of qualitative agreement is omnipresent in the statistical physics literature, well beyond the field of active matter. In many cases, taking interactions between particles into account in a quantitative way is very difficult, and mean-field approaches are developed, generically yielding qualitative agreement. For instance, the mean-field treatment of the Ising model cannot claim quantitative agreement with the nearest-neighbor Ising model in low dimension, but it can be considered as an important milestone in statistical physics.

In addition, the quantitative character of a theoretical approach can only be judged on its results, and not on its goal of being quantitative. We agree that the kinetic equations written in \cite{IHLE-DD,IHLE} may quantitatively describe the Vicsek model in the large speed limit where the molecular chaos is expected to hold (a property however difficult to prove, see \cite{Degond} for an attempt in this direction) and for arbitrary density, because multiparticle interactions are taken into account. However the kinetic equations are not in themselves a convenient macroscopic description so that quantitative tests are difficult at this level. When coming to hydrodynamic equations, that is the continuous equations describing the density and momentum (or polarity) fields, the claim of quantitative agreement made in \cite{IHLE-DD} has to be discussed.

\begin{figure}[t!]
\begin{center}
\includegraphics[width=0.8\columnwidth]{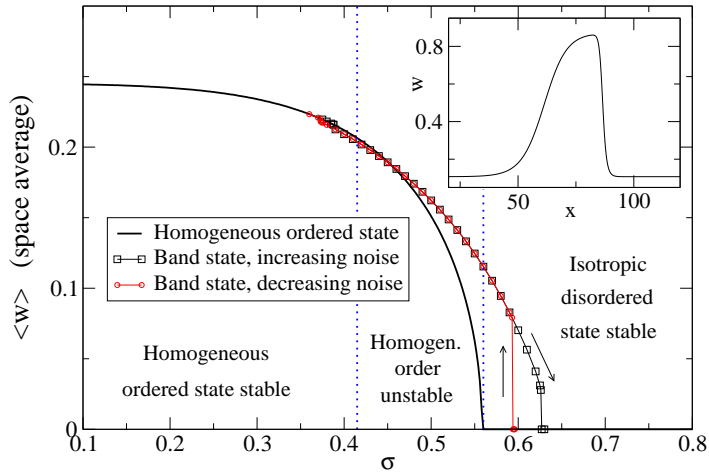}
\caption{(Color online) Illustration of the domain of existence of the travelling bands and of the hysteresis phenomenon, from numerical integration of the BDG hydrodynamic equations \cite{BDG}. The order parameter (average momentum $\langle w \rangle$) is plotted as a function of the noise amplitude $\sigma$ for a fixed value of the density. Vertical dotted lines indicate the limits of linear stability of homogeneous states, and the full line corresponds to the homogeneous ordered state, known analytically. Symbols correspond to travelling bands solutions, either when increasing (squares) or decreasing (circles) the noise (see also the arrows). An hysteresis behavior is observed at the transition. Note that the jump to the band solution (upward arrow) occurs at a noise value that depends on the details of the small perturbation around the homogeneous disordered state. Without any finite perturbation, the jump would occur at the linear instability threshold when decreasing the noise. Inset: illustration of the asymmetric shape of a band travelling to the right, obtained for $\sigma=0.5$.
}
\label{diag-band}
\end{center}
\end{figure}

First, it is known in the Vicsek model that close to the onset of order, the homogeneous ordered state is unstable, leading to travelling ordered bands \cite{CHATE}. The hydrodynamic equations derived in \cite{IHLE-DD,IHLE} indeed show this instability of the homogeneous ordered state, but fail to describe the smooth non-linear regime of travelling bands. Instead, their numerical integration leads to the appearance of singularities, meaning that the equations are ill-behaved. In contrast, the equations derived in \cite{BDG} are well-behaved and qualitatively describe the travelling band regime. Note that bands with much more pronounced asymmetry than that presented in \cite{BDG} can also be obtained \cite{ANTON_PHD}, as shown in the inset of Fig.~\ref{diag-band}. An illustration of the existence domain of bands and of the hysteresis phenomenon is provided in Fig.~\ref{diag-band}.

Second, the fact that Ihle's hydrodynamic equations do not account for bands has an important consequence on the characterization of the transition line between ordered and disordered states. The transition in the Vicsek model is known to be discontinuous and precisely related to the appearance of ordered bands. At the hydrodynamic level, the transition thus does not correspond to the linear instability threshold of the disordered state, but rather corresponds to the (upper) limit of the domain of existence of the ordered bands. As the hydrodynamic equations derived in \cite{IHLE-DD,IHLE} do not describe the ordered bands, they cannot describe the transition line in a quantitative way.

Finally, we note that the scaling of the density used in \cite{IHLE-DD,IHLE} with the deviation $\epsilon$ from the linear instability threshold, namely $\rho \sim \epsilon^0$, is different from the one we have used in our recent publications, under the name of Boltzmann-Ginzburg-Landau approach \cite{RODS-KINETIC,ACTIVE-NEMA}. There, expanding around the spatially homogeneous disordered state,
we have argued in favor of a scaling $\rho-\rho_0 \sim \epsilon$, where $\rho_0$ is the mean density (see also \cite{BGL-DD} for a more detailed discussion of this point). This change in scaling has important consequences. Applying the scaling relation $\rho-\rho_0 \sim \epsilon$ to the kinetic theory of \cite{IHLE-DD,IHLE} when deriving hydrodynamic equations, one gets back an equation with exactly the same terms as the BDG equation, apart from the values of the coefficients. The presence of additional terms in the hydrodynamic equations, under the scaling assumption $\rho \sim \epsilon^0$, may thus be responsible for the numerical instability of these equations, as described above.

\medskip
\noindent
{\bf Validity of the BDG Boltzmann equation.}
In \cite{IHLE-DD}, Ihle also provides a critical assessment of the validity of the BDG Boltzmann equation \cite{BDG} to describe the low density regime of the Vicsek model. While we technically agree with most of the results presented (see also our discussion in \cite{BGL-DD}), we do not share all the conclusions reached in \cite{IHLE-DD}. In particular, we never claimed that the Boltzmann equation introduced in \cite{BDG} was a low density description of the Vicsek model. Instead, it was made clear in these publications that the microscopic model considered was a continuous time variant of the Vicsek model, in which only binary interactions occur. Contrary to the implicit interpretation of \cite{IHLE-DD}, this model is not the limit of the Vicsek model for low density and vanishing time step, but a different model belonging to the same class.
In this continuous-time model (which, we acknowledge, was not described in full details in \cite{BDG}), binary alignment interactions are supposed to occur once per collision, exactly when the two particles reach the interaction range. Contrary to the Vicsek model, there is no repeated interactions during the time laps when the distance between particles is lower than the interaction range. 
Hence the arguments given in \cite{IHLE-DD}, although technically correct, do not rule out the correctness of the Boltzmann approach for the microscopic model we considered.
Interestingly, binary collisions seem nevertheless to dominate interactions in low density systems when positional diffusion (isotropic or anisotropic) is present, either alone \cite{ACTIVE-NEMA} or alongside drift \cite{BGL-DD}.
Moreover, we are also not interested in reproducing the zero and infinite speed limits, as they are clearly singular and correspond, respectively, to equilibrium dynamics and to a network with totally random rewiring \cite{Comment}. 

Finally, let us emphasize that even though the quantitative validity of the BDG Boltzmann equation is not ensured a priori, the hydrodynamic equations derived from it correctly describe, at a qualitative level, the transition scenario (instability of the homogeneous ordered state leading to travelling ordered bands, discontinuous transition due the existence of bands at density below the linear instability threshold of the disordered state). We consider this as an important a posteriori test of the qualitative validity of our approach which, among other things, correctly describes the qualitative dependence of the transport coefficients on the hydrodynamic fields.



\begin{thebibliography}{99}

\bibitem{IHLE-DD}
T. Ihle, {\it Towards a quantitative kinetic theory of polar active matter}, arXiv:1401.8056, to appear in Eur. Phys. J. Specials Topics Discussion and Debates (2014).

\bibitem{IHLE}
T. Ihle, Phys. Rev. E {\bf 83}, 030901 (2011).

\bibitem{VICSEK} 
T. Vicsek {\it et al.}, Phys. Rev. Lett. {\bf 75}, 1226 (1995).

\bibitem{CHATE}
G. Gr\'egoire and H. Chat\'e, Phys. Rev. Lett. {\bf 92}, 025702 (2004);
H. Chat\'e {\it et al.}, Phys. Rev. E {\bf 77}, 046113 (2008).

\bibitem{BDG}
E. Bertin, M. Droz, and G. Gr\'egoire,
Phys. Rev. E {\bf 74}, 022101 (2006); J. Phys. A {\bf 42}, 445001 (2009).

\bibitem{RODS-KINETIC}
A. Peshkov, I.~S. Aranson, E. Bertin, H. Chat\'e, F. Ginelli, Phys. Rev. Lett. {\bf 109}, 268701 (2012).

\bibitem{ACTIVE-NEMA}
E. Bertin, H. Chat\'e, F. Ginelli, S. Mishra, A. Peshkov, S. Ramaswamy, New J. Phys. {\bf 15}, 085032 (2013).

\bibitem{TOPOKINETIC}
A. Peshkov, S. Ngo, E. Bertin, H. Chat\'e, F. Ginelli, Phys. Rev. Lett. {\bf 109}, 098101 (2012).

\bibitem{IHLE-TOPO}
Y.-L. Chou, R. Wolfe, T. Ihle, Phys. Rev. E {\bf 86}, 021120 (2012).

\bibitem{BGL-DD}
A. Peshkov, E. Bertin, F. Ginelli, H. Chat\'e, {\it Boltzmann-Ginzburg-Landau approach for continuous descriptions of generic Vicsek-like models}, arXiv:1404.3275,
to appear in Eur. Phys. J. Specials Topics Discussion and Debates (2014).

\bibitem{ANTON_PHD}
A. Peshkov, {\em Boltzmann-Ginzburg-Landau approach for simple models of active matter}, Ph.D thesis, Universit\'e Pierre et Marie Curie, Paris (2013).

\bibitem{MCM-polar}
S. Mishra, A. Baskaran, and M.C. Marchetti, Phys. Rev. E {\bf 81}, 061916 (2010).

\bibitem{Gautrais}
J. Gautrais, F. Ginelli, R. Fournier, S. Blanco, M. Soria, H. Chat\'e, and G. Theraulaz, PLoS Comp. Biol. {\bf 8}, e1002678 (2012).

\bibitem{Weber-VPD}
C.~A. Weber, T. Hanke, J. Deseigne, S. L\'eonard, O. Dauchot, E. Frey, and H. Chat\'e,
Phys. Rev. Lett. {\bf 110}, 208001 (2013).

\bibitem{Degond}
E. Carlen, R. Chatelin, P. Degond, B. Wennberg, Physica D {\bf 260}, 90 (2013);
E. Carlen, P. Degond, B. Wennberg, Mathematical Models \& Methods in
Applied Sciences, {\bf 23}, 1339 (2013). 

\bibitem{Comment}
H. Chat\'e, F. Ginelli, and G. Gr\'egoire, Phys. Rev. Lett. {\bf 99}, 229601 (2007).

\end{thebibliography}
\end{document}